\documentclass[aps,prl,showpacs,twocolumn,amsmath,amssymb,
superscriptaddress]{revtex4}

\usepackage{graphicx}
\usepackage{hyperref}
\begin{document}
\title{Terahertz Bloch oscillator with a modulated bias}
\author{Timo Hyart}
\affiliation{Department of Physical
Sciences, P. O. Box 3000, FI-90014 University of Oulu, Finland}
\affiliation{Department of Physics, Loughborough University LE11 3TU,
United Kingdom}

\author{Natalia V. Alexeeva}
\affiliation{Micro- and Nanotechnology Center, FIN-90014 University of Oulu, Finland}

\author{Jussi Mattas}
\affiliation{Department of Physical Sciences,  P. O. Box 3000, FI-90014 University of Oulu, Finland}

\author{Kirill N. Alekseev}
\affiliation{Department of Physics, Loughborough University LE11 3TU,
United Kingdom}
\affiliation{Department of Physical
Sciences, P. O. Box 3000, FI-90014 University of Oulu, Finland}

\begin{abstract}
Electrons performing Bloch oscillations in an energy band of a
dc-biased superlattice in the presence of 
weak dissipation can potentially generate THz fields at room
temperature. The realization of such Bloch oscillator is a
long-standing problem due to the instability of a homogeneous electric field in conditions
of negative differential conductivity. We establish the theoretical
feasibility of stable THz  gain in a long superlattice device in
which the bias is quasistatically modulated by microwave fields. The
modulation waveforms must have at least two harmonics in their spectra.
\end{abstract}

\maketitle

\begin{figure}
\includegraphics[width=0.49 \columnwidth]{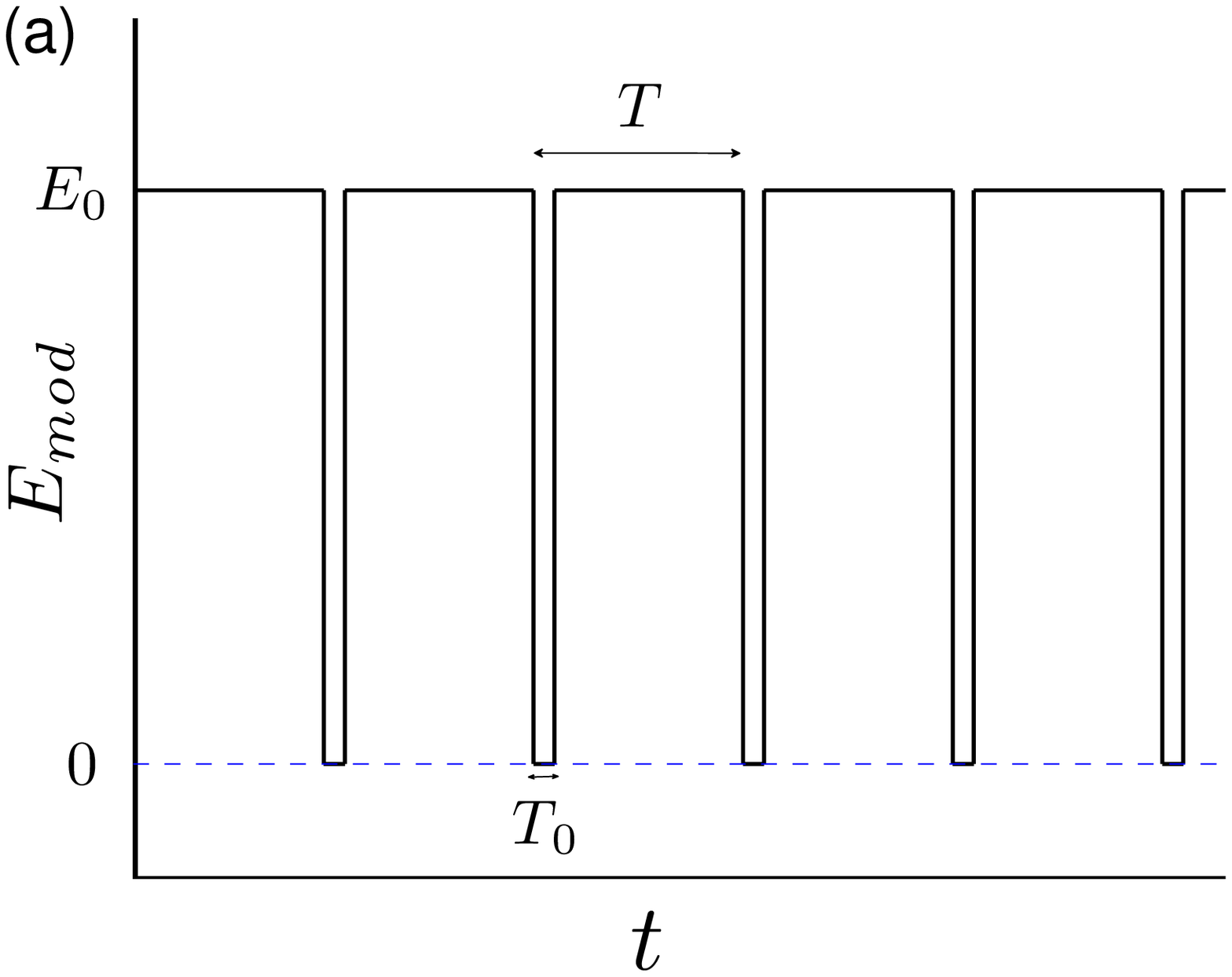}
\includegraphics[width=0.49 \columnwidth]{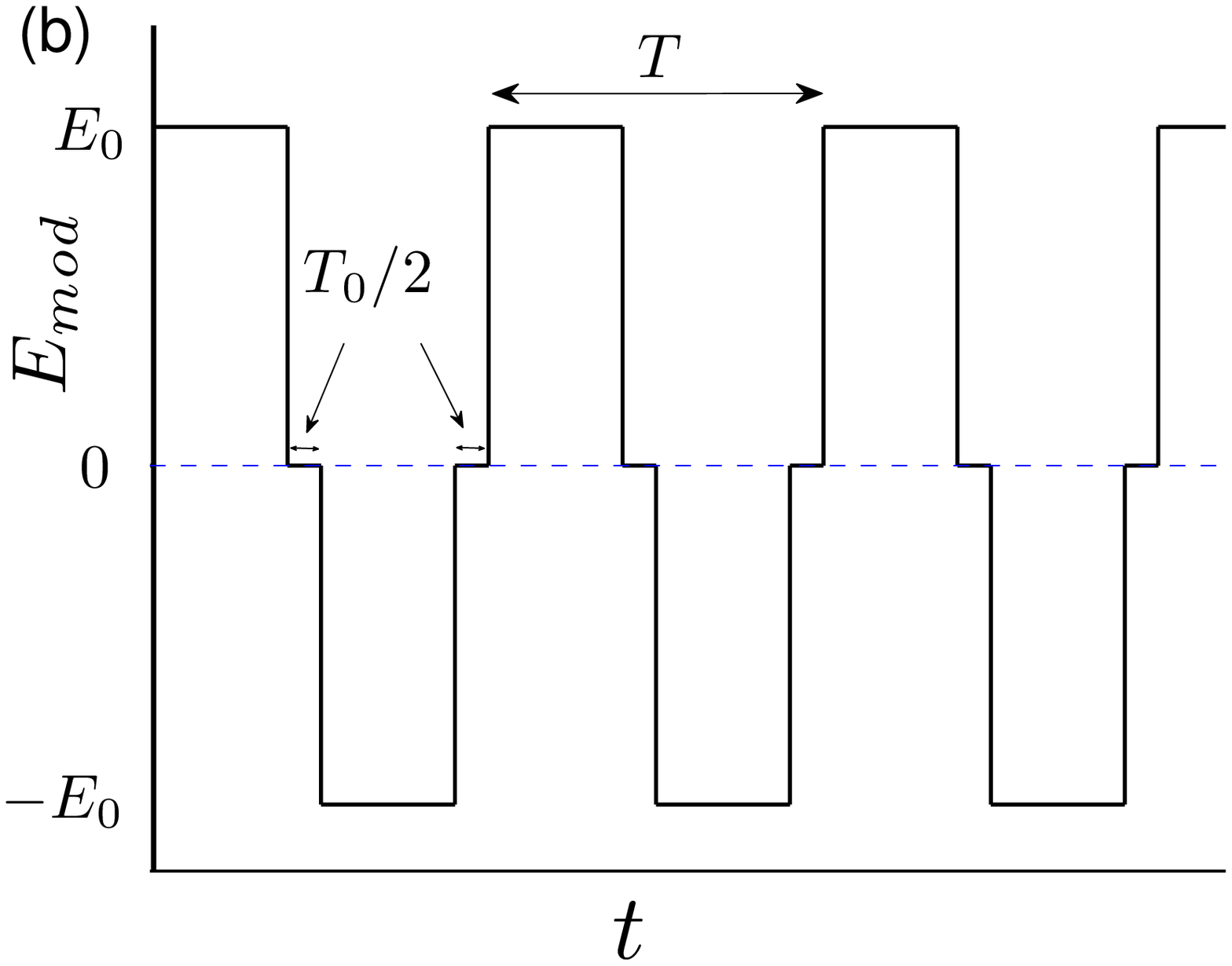}
\includegraphics[width=0.49 \columnwidth]{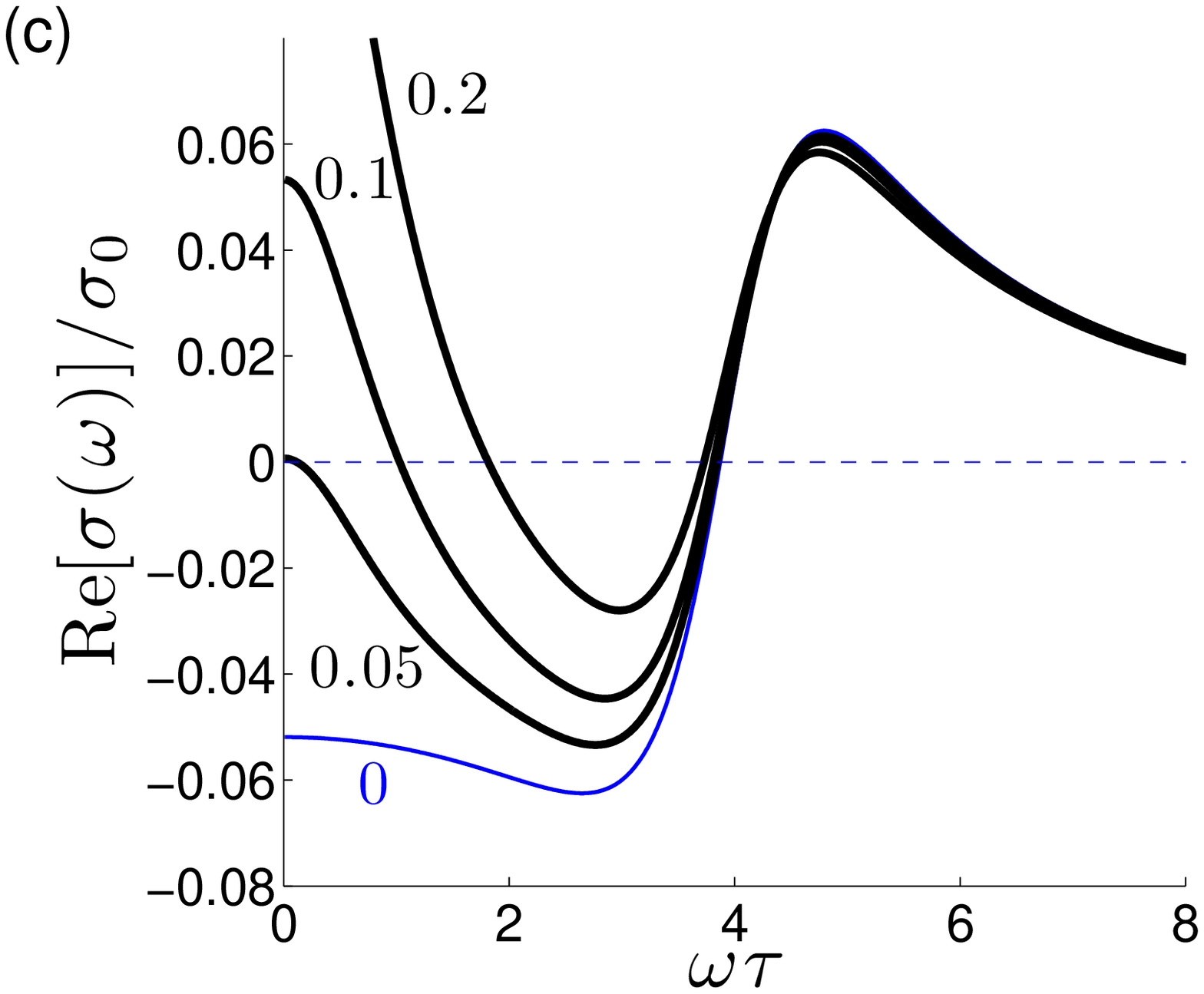}
\includegraphics[width=0.49 \columnwidth]{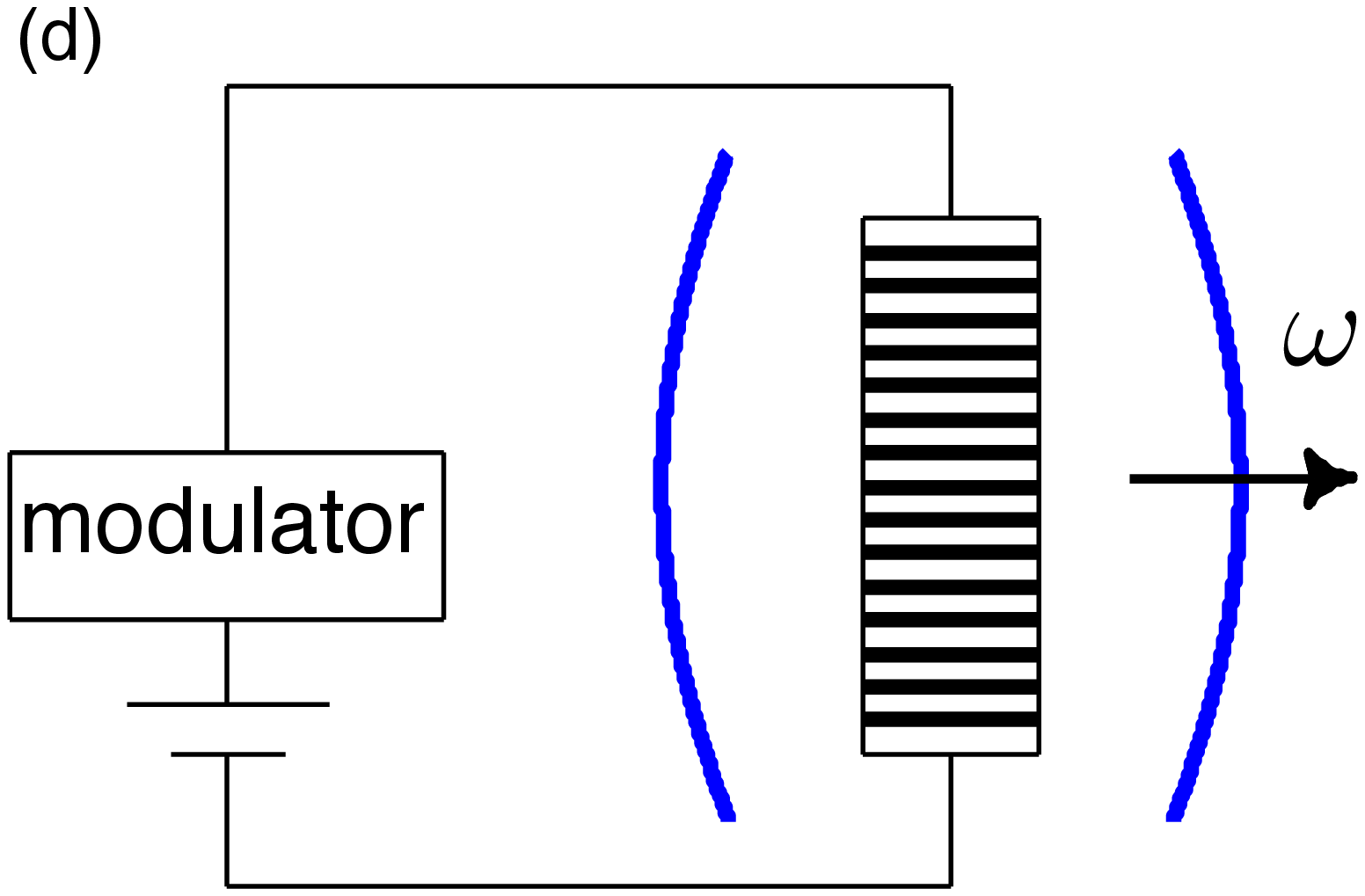}
\caption{ (a), (b) Two different waveforms of bias modulation
$E_{mod}(t)$ which result in the same dynamical conductivity of SL
[Eq.~(\ref{gainpulse})]. (c) Gain profiles of SL 
calculated using Eq.~(\ref{gainpulse}) for $E_0=4 E_{cr}$ and
$T_0/T=:0,0.05, 0.1, 0.2$. (d) Sketch of the Bloch oscillator with a
modulated bias.} \label{fieldprof}
\end{figure}

Electrically biased semiconductor superlattice (SL), which
operates in conditions of a single miniband transport regime,
exhibits a static negative differential conductivity (NDC) due to
Bloch oscillations \cite{Esaki}. Assuming a homogeneous electric
field inside the nanostructure, the semiclassical theory predicts
that SL in the  NDC state can provide small-signal gain for all
frequencies below the Bloch frequency \cite{KSS}. If miniband electrons can perform
relatively many cycles of Bloch oscillations between scattering
events, the Bloch gain profile is shaped as a familiar dispersion
curve (Fig.~\ref{fieldprof}c, blue curve). In general, the
dispersive gain profile is a quantum dissipative phenomena which is
not limited to the Bloch oscillating electrons. Similar profiles
with a resonant crossover from loss to gain have been observed in
optical transitions of quantum cascade lasers \cite{qcl-bloch} and
microwave response of Josephson junctions
\cite{jj-bloch}. Importantly, estimates for typical SLs
predict a significant THz gain even at room temperature for
frequencies in the vicinity of Bloch gain maximum. Thus, the
principal scheme of CW Bloch oscillator would consist of a biased SL
which is placed in a high-$Q$ cavity tuned to a desirable THz
frequency. However, even for moderate carrier densities a state with
NDC is unstable against a development of space-charge instability
which eventually results in the formation of high-field domains inside 
a long sample \cite{Ridley63-ignatov87}. The formation of electric
domains, which breaks the assumption of a homogeneous electric field
used in \cite{KSS}, ultimately represents one of the major obstacles
to the realization of Bloch oscillator.
\par
In short SLs the formation of electric domains can be suppressed. However, since
a short sample can emit only a  weak electromagnetic power a stack
of such SLs is required to observe the Bloch gain \cite{savvidis}.
In our research we focus on the feasibility of Bloch gain in a long
SL. To circumvent the electric instability in a long SL it is
necessary to modify the unstable Bloch gain profile in such a way
that the sample exhibits a positive dynamical conductivity at low
frequencies, including dc, whereas the high-frequency conductivity
is still negative giving rise to the desired amplification
\cite{macgroddy, wacker_q-deriv}. To reach this goal it was
suggested to use SLs with narrow gaps, where additionally to the
intraband Bloch oscillations the interband Zener tunneling  becomes
important \cite{andronov}. In a recent publication \cite{hyart08},
we theoretically showed that the desirable modification of Bloch
gain profile still can be achieved within the standard single
miniband transport regime. 
We utilized additional peaks in the voltage-current (VI) characteristics of SL, which arise under the action
of a strong ac field if the Bloch frequency is close to the ac field frequency or its harmonics \cite{unterrainer}. 
However, the need for the use of a strong THz field as ac pump limits the applied
aspect of this suggestion.
\par
In this Letter, we show that the unstable Bloch gain profile can be
made stable by means of a low-frequency modulation of
dc bias applied to a long SL. At least two harmonics with a proper phase difference are necessary
to be present in the modulation spectrum in order to suppress the
destructive low-frequency instability. The degree to which the large
magnitude of THz gain near the Bloch resonance may be preserved here
depends on the modulation waveform. In this respect we find the
pulse modulation optimal. Since the operation of Bloch oscillator
requires that during at least some portion of the modulation period
SL is biased to the NDC state, an accumulation of charge can arise. The
requirement that the charge accumulation is weak
imposes a lower limit on the modulation frequency, which, however, still belongs to the microwave
frequency domain. Finally, our numerical simulations show that the
effect of stable THz gain persists if boundary
conditions in the SL device are taken into account.
\par
We consider a SL under the action of an electric field
$E(t)=E_{mod}(t)+E_\delta \cos\omega t$, where $E_\delta \cos\omega t$ is a weak probe field
with the frequency $\omega$ fixed by an
external circuit (external cavity) and $E_{mod}(t)$ is a periodic
modulation of the bias (Fig.~\ref{fieldprof} (d)). We suppose that the modulation frequencies are incommensurate with $\omega$.
To understand the main idea of our suggestion let us first
consider $E_{mod}(t)$ in the forms of pulses as shown in
Fig.~\ref{fieldprof} (a) and (b). For both the asymmetric (a) and
symmetric (b) waveforms the amplitude of modulation $E_0$ is such
that during each period $T$ SL spends most of the time $T-T_0$ in
the NDC region of the VI characteristic and only for a short
fraction of the period $T_0/T \ll 1$ the bias is switched off. To
calculate the gain profile one needs to know 
the real part of the dynamical conductivity $\sigma_r(\omega)\equiv
{\rm Re}[\sigma(\omega)]$. Gain corresponds to $\sigma_r(\omega)<
0$. We assume that the switching is faster than both $T_0$ and
$T-T_0$ and transients after the switching can be neglected $T-T_0,
T_0 \gg\tau$ ($\tau$ is the intraband relaxation time)
\cite{lisauskas08}. We find for the both sequences of pulses
\begin{equation}
\sigma_r(\omega) = \left(1-\frac{T_0}{T}\right) \sigma_r^{KSS}
(\omega)+\frac{T_0}{T} \sigma_r^{free} (\omega),
\label{gainpulse}
\end{equation}
where
\begin{eqnarray}
\sigma^{KSS}_r(\omega)&=&\frac{j_{dc}(eE_{0}d+\hbar
\omega)-j_{dc}(eE_{0} d-\hbar \omega)}{2 \hbar \omega}ed, \label{eq:KSS}\\
\sigma^{free}_r(\omega) &=&\sigma_0 \frac{1}{1+\omega^2 \tau^2} \label{eq:free}
\end{eqnarray}
are the dynamical conductivity of a biased SL
and free-electron absorption in an unbiased SL, respectively. Here
$j_{dc}(e E_{dc} d)$ is the Esaki-Tsu characteristic
\cite{Esaki,wackerrew}
\begin{equation}
\label{eq:ET} j_{dc}(e E_{dc} d)=j_p \frac{2 e E_{dc}
d \tau/\hbar}{1+(e E_{dc} d \tau/\hbar)^2},\quad
\end{equation}
$$
j_p=\frac{e N \Delta d}{4 \hbar}
\frac{I_1(\frac{\Delta}{2k_BT_e})}{I_0(\frac{\Delta}{2k_BT_e})}
$$
is the peak current corresponding to the critical field 
$E_{cr}=\hbar/ed\tau$, $I_k(x)$ ($k=0,1$) are the modified
Bessel functions, $N$ is the density of electrons in the first
miniband of the width $\Delta$, $d$ is the SL period, $T_e$ is the
temperature and $\sigma_0=2j_{p}/E_{cr}$ is the Drude conductivity of SL.
Note that the finite difference form \cite{tucker79} given by
Eq.~(\ref{eq:KSS}) is equivalent \cite{wackerrew} to the original
Ktitorov \textit{et al.} (KSS) formula \cite{KSS} for a single relaxation time. 
\par
In the limit $T_0/T \to 0$, Eq.~(\ref{gainpulse}) describes the
usual Bloch gain profile (thin blue line in Fig.~\ref{fieldprof}
(c)). With an increase of $T_0/T$ SL spends more time in an unbiased
state, therefore the gain at low frequencies decreases and
eventually the dynamical conductivity becomes positive, whereas the
high-frequency gain is still preserved as shown in Fig.~1 (c). The
physical origin of such behavior is quite intuitive: At high
frequencies around $\omega \approx \omega_B-\tau^{-1}$
($\omega_B\equiv eE_{0}d/\hbar \gg \tau^{-1}$ is the Bloch
frequency), the Bloch gain $\sigma_r^{KSS}(\omega)\approx
-\sigma_0/4\omega\tau$ dominates the free-electron absorption
$\sigma_r^{free}(\omega)\approx \sigma_0/ \omega^2\tau^2$. In
contrast, at low frequencies $\omega \tau \ll 1$, the free-electron
absorption $\sigma_r^{free}\approx \sigma_0$ overcomes the Bloch
gain $\sigma_r^{KSS}\approx -\sigma_0/\omega_B^2 \tau^2$.
\par
The dynamic conductivity expressions Eqs.~(\ref{gainpulse})-(\ref{eq:free}) directly follow from 
the general formula describing the dynamic conductivity of SL in the case of arbitrary slow (quasistatic) modulation
\begin{equation}
\label{Aictri}
\sigma_r(\omega)=\frac{j_{dc}^{mod}(eE_{dc}d+\hbar\omega)-j_{dc}^{mod}(eE_{dc}d-\hbar\omega)}{2\hbar\omega}
e d,
\end{equation}
where $j_{dc}^{mod}$ is the dc component of the current in SL
\begin{equation}
\label{tasavirtatri} j_{dc}^{mod} (eE_{dc}d)=\frac{1}{T}\int_0^T
j_{dc}\left[eE_{mod}(t)d\right] \ dt
\end{equation}
that adiabatically follows to the time-dependent variations
$E_{mod}(t)$ via the Esaki-Tsu characteristic (\ref{eq:ET}). 
These formulas can be derived using two different techniques.
In the first method, we employed the standard semiclassical approach based
on an exact solution of Boltzmann transport equation for a single
tight-binding miniband and a constant relaxation time
\cite{wackerrew}. Next, we also found that the same dependence of
$\sigma(\omega)$ on the shape of VI characteristic
can be obtained in Tucker theory of quantum tunneling in
the presence of ac electric fields \cite{tucker79}. Both approaches
result in Eq.~(\ref{Aictri}) with a cumbersome expression for the
$j_{dc}^{mod}$, which can be further reduced \cite{shorokhov} to the
simple form of Eq.~(\ref{tasavirtatri}) in the limit of low-frequency modulations.
\par
For the pulse modulation the spectrum of $E_{mod}(t)$
consists of many harmonics of the basic frequency $\Omega(=2\pi/T)$ and, 
as we have seen, the gain profile has the desirable shape with
$\sigma_r(\omega\to 0)>0$. Importantly, in the 
case of  monochromatic quasistatic field no high-frequency gain exists 
if $\sigma_r(\omega\to 0)>0$ \cite{shorokhov}. 
Therefore, we conclude that the \textit{monochromatic} low-frequency modulation cannot provide the 
required stabilization of the space-charge instability in the Bloch oscillator.
Now we are going to show that a bichromatic modulation can
already mimic the kind of fast bias switching necessary for the stabilization. We consider a modulation 
\begin{equation}
\label{E-trichrom}
E_{mod}(t)=E_{dc}+E_{1}\cos{\Omega t}+E_{2}\cos(n\Omega t+\phi),
\end{equation}
where $n$ is an integer and $\Omega\tau\ll 1$.
\begin{figure}
\includegraphics[width=0.49\columnwidth]{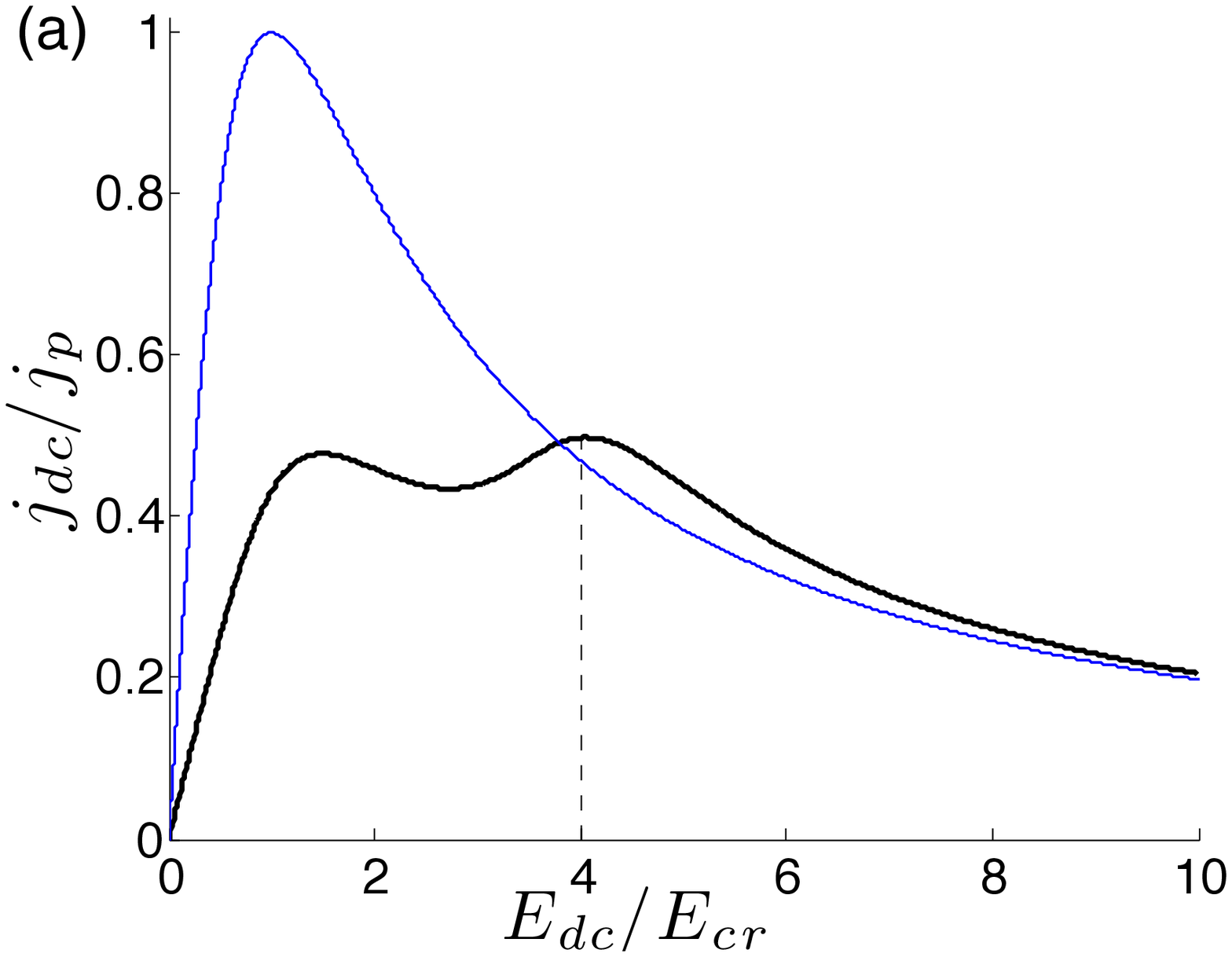}
\includegraphics[width=0.49\columnwidth]{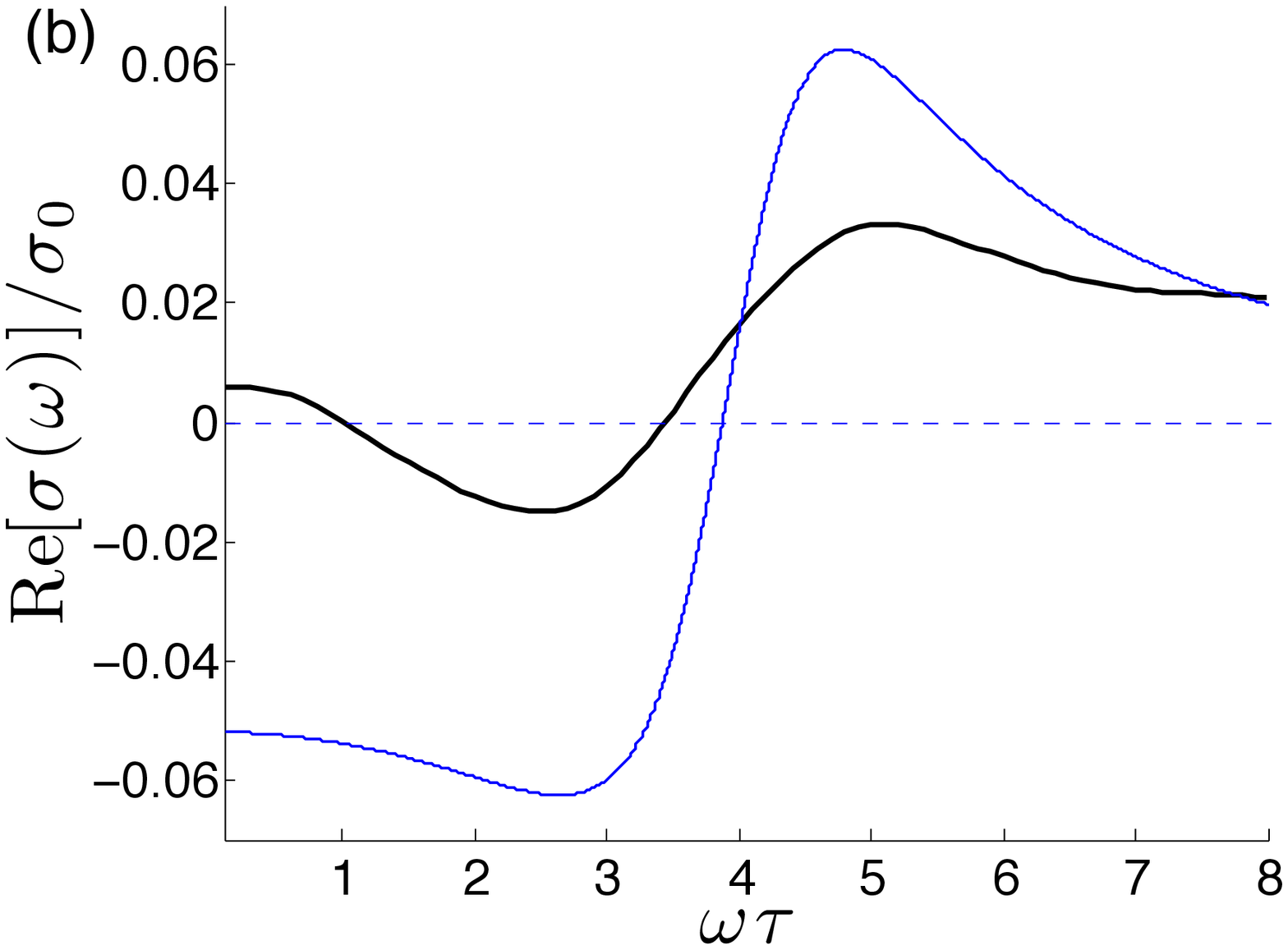}
\caption{(a) 
Esaki-Tsu characteristic (thin blue line)
[Eq.~(\ref{eq:ET})] and the characteristic of SL  modified by the
bichromatic field  [Eq.~(\ref{tasavirtatri})] (thick black line) for
$E_1=2.1 E_{cr}$, $E_2=1.7 E_{cr}$, $\Omega\tau=0.05$, $n=3$,
$\phi=0$. (b) Dynamical conductivity  [Eq.~(\ref{Aictri})] as a
function of $\omega$ (thick black) for the same bichromatic modulation and
$E_{dc}=4E_{cr}$. The thin blue curve shows the usual Bloch gain profile [Eq.~(\ref{eq:KSS})].}
\label{bichroBloch}
\end{figure}
Fig.~\ref{bichroBloch} demonstrates the VI characteristic and gain
profile for the modulation with moderate field strengths $E_{1,2}$.
The amplitude $E_2$ of the harmonic component and the relative phase
$\phi$ are chosen to square up the field (\ref{E-trichrom}). For
typical scattering time $\tau=200 \ \rm{fs}$ the modulation
frequency used in the figure is $\Omega/2\pi \approx 40 \ \rm{GHz}$.
This $E_{mod}(t)$ creates, following Eq.~(\ref{tasavirtatri}), an additional peak in the VI characteristic of SL (Fig.~\ref{bichroBloch}(a)), which resembles 
the resonant structures in VI characteristic induced by a monochromatic
THz field \cite{unterrainer}. This additional peak can be utilized
in a similar way as in the earlier theoretical suggestion
\cite{hyart08} to achieve THz gain at a stable operation point.
Namely, we put the working point at the part of the peak with a
positive slope and then apply the difference formula (\ref{Aictri}).
The result is the dispersive gain profile with a positive dynamical
conductivity at low frequencies as shown in
Fig.~\ref{bichroBloch}(b). In contrast to the earlier work
\cite{hyart08}, for the suppression of the electric instability in
SL we use a bichromatic microwave field \cite{remark1} instead of a
monochromatic THz field. Importantly, the shape of the VI
characteristic and magnitude of gain can be additionally controlled
by a variation of the relative phase $\phi$. The magnitude of THz
gain at bichromatic modulation is a bit lower in comparison with the
case of pulse modulation (\textit{cf.} Figs.~\ref{bichroBloch}(b)
and \ref{fieldprof}(c)).
\par
Apart from $\sigma_r(\omega\to 0)>0$, the conditions for
space-charge control in microwave driven SL with NDC include also
the requirement $\Omega \tau_d > 1$ \cite{alekseev06}, where
$\tau_d$ is a characteristic time of domain formation. It means that an accumulation of charge during 
the period of modulation $T$ is always weak \cite{copeland,alekseev06}. If the conditions $\Omega \tau_d > 1$
and $\sigma_r(\omega\to 0)>0$ are satisfied, VI characteristic of SL
is modified by strong microwave fields according to the quasistatic
theory prediction with an assumption of spatially homogeneous
fields, \textit{i.e.} following Eq.~(\ref{tasavirtatri}). Experimental observation of such modifications in the case of
monochromatic microwave field has been reported in
\cite{schomburg96}. The time $\tau_d$ is of order of the dielectric
relaxation time $\epsilon \epsilon_0/|\sigma_{dc}(E_{dc})|$
\cite{zakharov60}, where $\sigma_{dc}(E_{dc})$ is the dc
differential conductivity, $\epsilon$ is the relative permittivity of SL material and
$\epsilon_0$ is the permittivity of vacuum.
For the Esaki-Tsu characteristic (Eq.~(\ref{eq:ET})) $\min [\sigma_{dc}(E_{dc})]=-\sigma_0/8$, 
and therefore we can present the condition of limited charge accumulation in the form
\begin{equation}
\label{LSA}
\omega^2_{pl} \tau^2  \lesssim 8 \Omega \tau,
\end{equation}
where $\omega_{pl}=\left( 2j_p e d/\hbar\epsilon_0\epsilon\right)^{1/2}$ is the miniband plasma
frequency \cite{epshtein77}. Since $\omega^2_{pl}\propto N\Delta$,
for given electron concentration $N$ and miniband width $\Delta$,
the inequality (\ref{LSA}) imposes a lower limit on the modulation
frequency $\Omega$. Alternatively, for a fixed value of the
modulation frequency $\Omega$, Eq.~(\ref{LSA}) requires the use of
SLs for which the product $N\Delta$ is less then some critical value.
\par
The absorption (gain) $\alpha$ in units $\rm{cm}^{-1}$ is related to the scaled dynamical conductivity as
\begin{equation}
\alpha=\omega_{pl}^2 \tau^2  \frac{\sqrt{\epsilon}}{c\tau}
\frac{\sigma_r(\omega)}{\sigma_0}, \label{alpha}
\end{equation}
where $c$ is the speed of light in vacuum. Comparing
Eqs.~(\ref{LSA}) and (\ref{alpha}) we see that the condition
(\ref{LSA}) imposes a restriction on the maximal value of $\alpha$.
For the parameters of bichromatic modulation considered in
Fig.~\ref{bichroBloch}, we estimate $\alpha\approx - 3.6 \ {\rm
cm}^{-1}$ at the gain resonance
$\omega\tau\approx 2.5$. The corresponding value $\omega^2_{pl}
\tau^2=0.4$ [Eq.~(\ref{LSA})] can be realized, for example, at room
temperature in SL with $d=6$ nm, $\Delta=60$ meV and $N=6\times
10^{15}$ cm$^{-3}$. The magnitude of
gain can be increased by using the rectangular wave modulations. In this case, for the same basic
modulation frequency $\Omega \tau=0.05$ we get $\alpha \approx -12.8
\ {\rm cm}^{-1}$ at THz gain resonance shown in
Fig.~\ref{fieldprof}(c). It requires the use of  $\simeq 10$ ps dc
pulses.  However, a proof-of-the-principle demonstration, perhaps in
the pump-probe configuration, can be performed with commercially
available electric pulses of the duration $\simeq 0.1$ ns. The
restriction (\ref{LSA}) can probably be relaxed in the case of
lateral surface SLs \cite{lateral} formed in two-dimensional
electron systems. Because the domain growth rate 
can be suppressed in 2D systems \cite{2D_domains}, we can expect reasonably large magnitudes of THz
gain even for a rather slow modulation of the bias.
\par
\begin{figure}
\includegraphics[width=0.49\columnwidth]{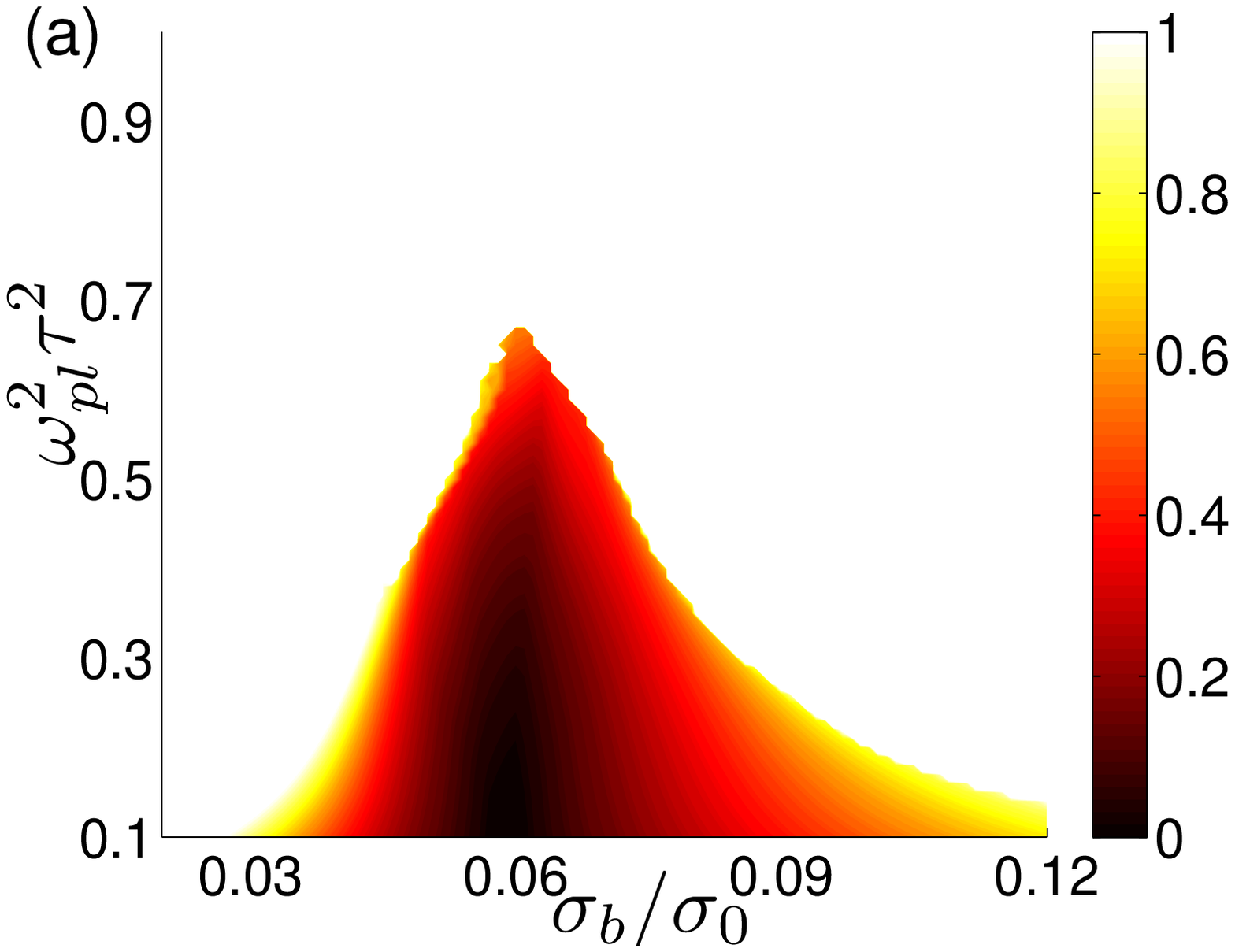}
\includegraphics[width=0.49\columnwidth]{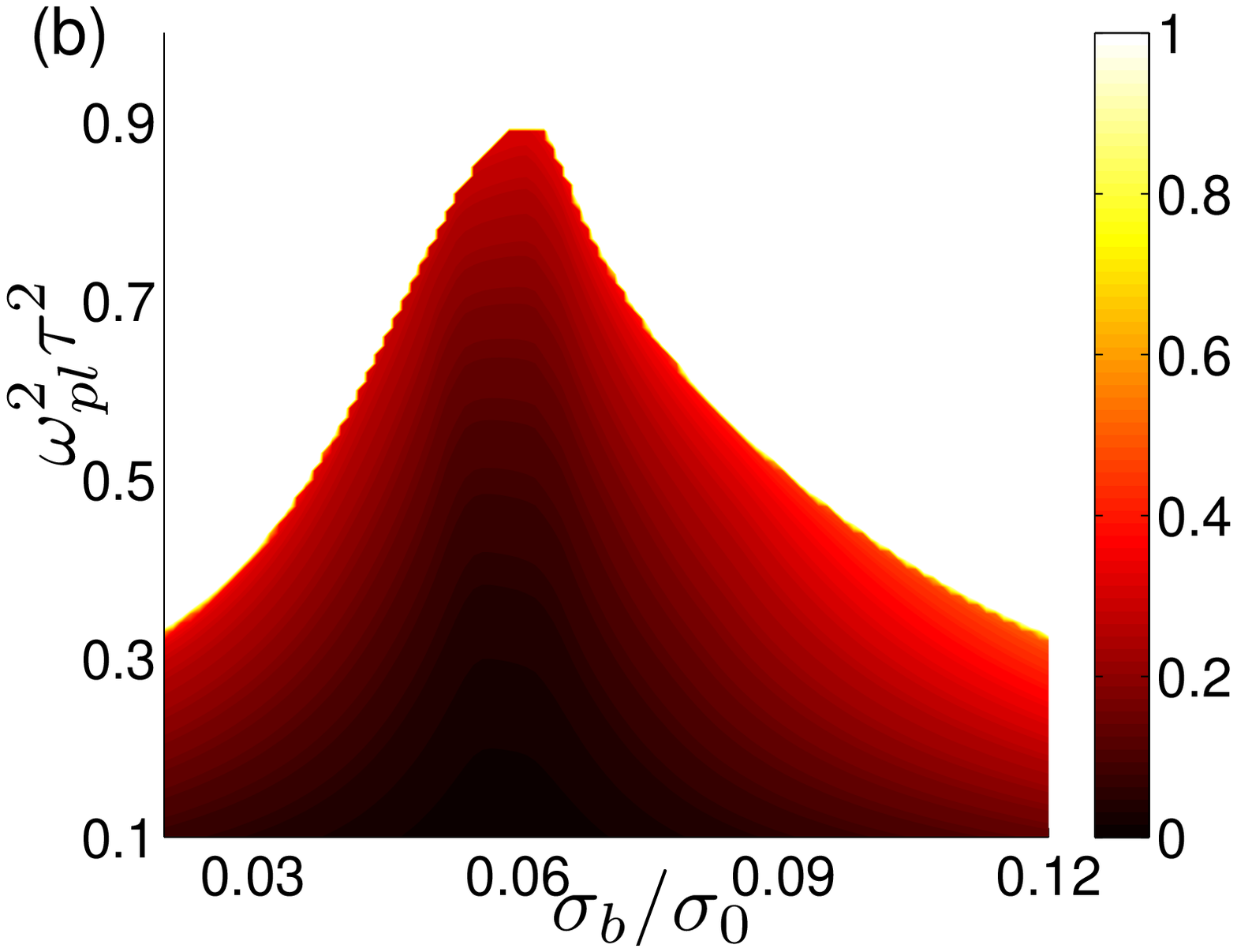}
\caption{Regions of a spatially homogeneous electric field (dark,
red online) in $\sigma_b$-$\omega^2_{pl}$ plane in the case of (a)
asymmetric and (b) symmetric pulse voltages (see
Fig.~\ref{fieldprof} (a) and (b), respectively) with $T_0/T=0.1$.
SL has the length $500 d$.} \label{homog_fig}
\end{figure}

Degree of homogeneity of the electric field inside
the SL also depends on the boundary conditions defined by the
attached electric contacts. Since the THz probe field is weak and
bias modulation is quasistatic, we can use the standard
drift-diffusion model \cite{alekseev06} with typical boundary conditions \cite{scheuerer}. 
SL with a nonlinear conductivity, which follows from the Esaki-Tsu dependence (\ref{eq:ET}), is placed
between two Ohmic buffer regions of the equal conductance $\sigma_b$
\cite{scheuerer}. The ac voltage across the device is periodically
varied as shown in Fig.~\ref{fieldprof}(a,b).
We numerically calculated the degree of homogeneity as 
$r=\langle\left[ E_{max}(t)-E_{min}(t) \right] /E_0\rangle_t$, where
$E_{max}(t)$ and $E_{min}(t)$ are the maximum and minimum of the electric field in the whole SL and
the time-averaging $\langle\ldots\rangle_t$ is performed over a long time interval of sustained field oscillations.
Fig.~\ref{homog_fig} shows the values of $r$ in the $\sigma_b$-$\omega^2_{pl}$ plane for a very long SL.
In the indicated dark regions of small $r$ values the electric field inside the SL can be
considered as spatially homogeneous. 
We also verified that for these parameters  spatially-averaged gain profiles \cite{lisauskas05}
are practically indistinguishable from those calculated earlier
(Fig.~\ref{fieldprof}(c)) within the homogeneous field approximation.
If the buffer conductance coincides with the SL conductance ($\sigma_b/\sigma_0\approx 0.06$ in Fig.~\ref{homog_fig}), 
the upper boundary of the homogeneous field region in $\omega^2_{pl}\tau^2$ is even a bit larger than expected from Eq.~(\ref{LSA}).
For mismatched conductances smaller values of $\omega^2_{pl}$ are required to achieve a
desirable homogeneity. Interestingly, in the case of symmetric modulation SL is less
sensitive to the boundary effects (\textit{cf.} (a) and (b) in
Fig.~\ref{homog_fig}). We attribute it to the fact that the
time-averaged current induced by the symmetric voltage is
zero both in the buffers and in SL independently on the value of
$\sigma_b$ and thus, on average, there should be 
no accumulation or depletion of charges at the boundaries. 
For shorter SLs we found that the difference between the effects of
symmetric and asymmetric waveforms becomes much less prominent and,
as expected, the field stays homogeneous for a much wider range of $\sigma_b$. 
We underline that an observation of the Bloch gain certainly requires a
gradual change of conductance between the  active SL region and the
metallic contacts.
\par
In summary, we have theoretically shown that a proper quasistatic modulation of the
bias in Bloch oscillator results in THz Bloch gain in conditions of positive differential conductivity.
Waveform of the modulation should include two distinct parts: Time
interval of almost constant bias responsible for the THz gain due to the excitation of Bloch oscillations and a virtually
unbiased interval during which the dominant free-carrier absorption
effectively suppresses the undesirable low-frequency instability.
Spectrum of the modulation must have at least two harmonics with a controlled difference of phases.
\par
Along with the semiconductor superlattices our results can be directly applied to such semi- and
superconducting microstructures as the dilute nitride alloys
\cite{patane} and small Josephson junctions \cite{small-jj}.
It has already been found that, despite of the very different
physical origin, the small Josephson junction demonstrate 
in fact the Esaki-Tsu VI characteristics and ac response 
similar to the single miniband superlattice \cite{small-jj}.

We thank Maxim Gorkunov and Alexey Shorokhov for
active collaboration and valuable advices at the initial stage of
the work, Amalia Patan\`{e} and Oleg Makarovsky for useful
discussion on the experimental feasibilities, Erkki Thuneberg and
Feo Kusmartsev for a constant encouragement of this activity. This
research was partially supported by V\"{a}is\"{a}l\"{a} Foundation, Emil Aaltonen Foundation,
Yliopiston Apteekki Foundation, Academy of Finland, University of Oulu, and AQDJJ Programme of European Science Foundation.

\end{document}